\documentclass[preprint,12pt]{elsarticle}
\usepackage{graphicx}
\usepackage{amssymb}
\usepackage{amsmath}
\usepackage{amsbsy} 
\usepackage{booktabs}
\usepackage{longtable}
\usepackage{epstopdf}
\usepackage{color}
\usepackage{subcaption}
\newcommand{\cfig}[1]{Fig[{\ref{#1}}]}
\newcommand{\ceq}[1]{eqn({\ref{#1}})}
\newcommand{\bee}{\begin{equation}}
\newcommand{\ee}{\end{equation}}

\newcounter{quotecount}
\newcommand{\MyQuote}[1]{\vspace{1cm}\hspace{1.1 cm}\refstepcounter{quotecount}%
     \parbox{10cm}{\em #1}\hspace*{2cm}(\arabic{quotecount})\\[1cm]}
\journal{IJTP}
\begin{document}
\begin{frontmatter}
\title{Superposition as a Relativistic Filter}
% Now published by IJTP DOI: 10.1007/s10773-017-3372-0
\author{G. N. Ord}

%\email[ \jobname \hspace{2mm} corresponding author   ]{gord@ryerson.ca}
\address{Department of Mathematics\\ Ryerson University\\Toronto Ont.}

%\maketitle
\begin{abstract}  
By associating a binary signal with the relativistic worldline of a particle,  a binary form of the phase of non-relativistic wavefunctions is naturally produced by time dilation.  An analog of superposition also appears as a Lorentz filtering process,  removing paths that are relativistically inequivalent. In a model that includes a stochastic component, the free-particle Schr\"{o}dinger equation emerges  from a completely relativistic context in which its origin {\em and function}  is known. The result establishes the fact  that the phase of wavefunctions in  Schr\"{o}dinger's equation and the attendant superposition principle may both be considered remnants of time dilation. This strongly argues that quantum mechanics has its origins in special relativity.
%\vspace{4cm}
\end{abstract}
\begin{keyword}
Quantum Mechanics, Special Relativity, Path Integrals
\PACS 03.65.-w \sep 03.65.Ud 
\sep 02.50.-r

\end{keyword}

\end{frontmatter}
\section{Introduction}

The objective of this paper is to explore {\em the origin and function} of the phase of wavefunctions in the solutions of Schr\"{o}dinger's equation, building on and clarifying some previous work \cite{OrdAnnals2009,OrdEmQm13}. As in emergent quantum mechanics\cite{Emerqm13}, the goal is to find an underpinning of Schr\"{o}dinger's equation, rather than an interpretation.

To a good approximation, the physics community is united in its agreement that the empirical  accuracy of non-relativistic quantum mechanics in its relevant domain exceeds that of any prior classical theory with the possible exception of relativity. In contrast, there has never been agreement on questions such as `What is a wavefunction?', `Is quantum mechanics complete?', `Is wavefunction collapse real?', `Are questions about interpretation of any importance?'  It is as if there is consensus on the grounds of empirical accuracy that wavefunctions are `the answer', but we do not quite have a precise formulation of the question.\footnote{ J. S. Bell, whose deep insights into the foundations of quantum mechanics have informed generations of physicists, lamented the lack of an `exact' theory underlying quantum mechanics. With incisive humour in his last publication\cite{JBell90a},  he labeled the current versions of quantum mechanics as good {\em For All Practical Purposes} (FAPP) in order to deflect criticism from those convinced of completeness  through familiarity with empirical accuracy. 
 This paper argues that while non-relativistic quantum mechanics {\em as a description} is good FAPP,    the origin of superposition and the roots of its strange behaviour are missing in the absence of relativity.}

To put the problem in context, compare the two equations:
\begin{equation}
\frac{\partial u}{\partial t}= D \frac{\partial^{2}u}{\partial x^{2}}\label{eqn:dif}
\end{equation}
and
\begin{equation}
\frac{\partial u}{\partial t}= i D \frac{\partial^{2}u}{\partial x^{2}}\label{eqn:shrod}
\end{equation}
where $D$ is a positive constant and $i$ is the unit imaginary.

The diffusion equation(\ref{eqn:dif}) occurs in a wide variety of contexts. The solutions may be written as probability density functions and derivations of the equation from elementary probability theory are well known.  The partial differential equation supports  a superposition principle that is expected both from the linearity of the  equation itself {\em and} from the probabilistic nature of the solutions in the context of classical statistical mechanics.  

By comparison, the linearity of Schr\"{o}dinger's equation(\ref{eqn:shrod}) dictates a superposition principle, but since wavefunctions are essentially {\em square roots} of probability densities, `quantum' superposition runs counter to  a classical expectation that, for example,  probabilities of {\em manifestly disjoint} events should add. 

The Young double slit experiment for electrons is a familiar example that displays this contrast well. That waves propagating through two slits should add seems natural enough until  the arrival of individual particles   at the detector screen are individually resolved and separated in time. The comparison with  experiment then shows that events corresponding to passage through one or the other slit {\em cannot be the disjoint events that would be expected for classical particles.} 

While the wavefunction solutions of Schr\"{o}dinger's equation and the associated superposition principle  provide a fundamental description of processes happening on atomic scales, questions remain as to what wavefunctions represent,  why they usurp the superposition principle from the probability density functions they represent and why Born's postulate connects wavefunctions to probabilities.

This paper argues that the {\em mechanism} of superposition of  Schr\"{o}dinger's equation originates in special relativity.   Since  Schr\"{o}dinger's equation has infinite signal velocity and is usually considered non-relativistic,   the sense in which it can have relativistic origins requires some explanation. 
  From a practical standpoint, % even in the foundations of quantum mechanics, 
  special relativity is conventionally ignored  in favour of explicitly non-relativistic mechanics provided characteristic velocities are much less than $c$. Its neglect in non-relativistic quantum mechanics is usually based on arguments along the following lines.
 
 {\centering \MyQuote{
Newtonian physics is obtained from relativistic mechanics by  judicious application of a small speed limit, frequently implemented by increasing the signal velocity $c\to \infty$ in relation to the characteristic speeds in the system. This limit, suitably applied, removes the physical aspects of length contraction and time dilation which are in any case negligible in systems where characteristic speeds are small.\\
  %Newtonian physics as an approximation to relativistic physics is a  set of relations in which negligible effects due to  time dilation and length contraction are removed from consideration. % on the ratio of any characteristic velocity $v$ to the speed of light $c$ (ie. $v/c$) are dropped. 
    Non-Relativistic Quantum Mechanics  represents the `quantization' of  such systems with the expectation that if relativistic effects are negligible in classical systems, they will remain so in quantum systems. NRQM is thus independent of physical manifestations of relativity and  relativistic quantum mechanics can effectively be regarded  as an {\em extension} of quantum mechanics to the relativistic domain. \label{quote:arg}
}
}

 Here,  the effectiveness of NRQM is not in question and for all practical purposes, the above argument is consistent with  the routine use of the Schr\"{o}dinger equation where characteristic speeds are small. However, if the objective is to extract quantum mechanics from a deeper level theory, the informal nature of the argument is suspect.
  
For example, if Nature is intrinsically discrete, then any route from a precise  discrete description   to the  Schr\"{o}dinger equation must involve at least two competing {\em approximations}. One approximation would involve the construction of a spacetime continuum, allowing wavefunctions to be defined on a continuous manifold.

 \noindent A second  would impose a restriction of characteristic velocities in relation to $c$ so as to suppress overt relativistic effects. However, quantum mechanics in general tells us that the two limits, involving both spacetime {\em and} momenta, cannot be independent. Such limits  are restricted by the uncertainty principle. In light of this, how and when limits are taken is of great importance and  this paper will emphasize two results that arise from an examination of competing limits, starting from a discrete model in which the worldline of a particle carries a binary signal.
 \begin{enumerate}
\item[A)]
 The phase of Schr\"{o}dinger wavefunctions is a manifestation of relativistic time dilation {\em given discrete time evolution at the Compton scale.} It  survives the  $c \to \infty$ limit in the transition from relativistic  mechanics to the Schr\"{o}dinger regime,  its relativistic origins {\em being hidden in the process.} From this perspective, canonical quantization from Newtonian mechanics replaces an aspect of time dilation {\em lost in the transition from classical relativistic to Newtonian physics.}
 
 \item[B)] Wavefunctions  occurring in this way operate as `Lorentz' filters, implementing a form of Lorentz invariance.   Superposition of wavefunctions takes precedence over superposition of probabilities in the quantum context because addition of  wavefunctions {\em preprocess} a signal to ensure that the ensemble of relevant alternatives for the system, from a probabilistic perspective, is consistent with relativity and the existence of a {\em single} worldline signal. The preprocessing effectively redefines what is meant by `mutually exclusive events' and Born's rule applies a probabilistic interpretation to a filtered ensemble of paths. 
 \end{enumerate}
  
 Neither A) nor B) is immediately obvious from non-relativistic quantum mechanics which effectively takes a continuum limit {\em prior} to considering the effect of time dilation. It is only by actually taking appropriate limits, starting from discrete processes, that A) and B) above become apparent. 
  
  The following article approaches the relevant limits  in two ways. The first section displays the `smoking gun'  that implicates  special relativity as the source of quantum phase. The Feynman propagator is compared to a binary signal of a {\em classical} relativistic clock running at the Compton frequency. At small velocities and fixed $t$, the signals are exactly synchronized, suggesting the possibility that the propagator is the binary signal `softened' by statistical averaging.  However, {\em the binary signal}  has a function not immediately visible in the propagator. It acts to filter available paths into an ensemble  with a  form of Lorentz invariance that is consistent with restrictions to images of a {\em worldline signal}.  
      
The second section explores a specific stochastic model that implements the picture sketched in the first section. The model starts with a simple binary clock on a lattice in a two dimensional spacetime. In the `diffusive' continuum limit, the Lorentz invariance may be maintained or ignored. In the former case one obtains the Schr\"{o}dinger equation directly, in the latter the diffusion equation.  The distinction between the two is relativistic from both the mathematical perspective of the limit taken and from the physics it represents. 
 \section{The Clock Model}

 One feature that is shared by special relativity and pre-relativistic mechanics is the concept of the worldline of a particle. There are of course differences. In the relativistic case, the slope of the worldline is limited by $c$, and the two versions transform differently between coordinate systems. However, in both cases the resulting curve considered {\em as a signal in spacetime} is a constant function, or delta-function, and neither identifies the mass of a particle, or any  other intrinsic feature. The worldline is simply a continuous curve, the points in the curve being considered {\em events} in a spacetime, indicating  persistent  existence and movement. 
 
\begin{figure}[htbp] %  figure placement: here, top, bottom, or page
    \centering
    \includegraphics[width=2in]{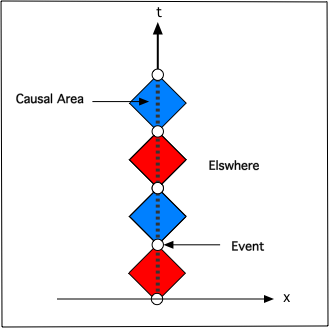} 
    \caption{Three events in a C-clock model. Time is vertical in the sketch, space is horizontal and the speed of light is 1. The worldline, considered as a linear interpolant between events at the Compton frequency of the particle, carries a binary signal that switches at events. Between each event is a causal spacetime area in which spacetime points are time-like connected to both events. Successive spacetime areas are distinguished in the signal associated with the worldline giving two colours in the sketch.  }
    \label{fig:incip}
 \end{figure}

 In the clock model under consideration (subsequently called a Clock-particle or C-particle), we alter this by distinguishing a {\em periodic} sequence of points on the worldline to act as  an event{ \em sequence}. Each  event toggles a binary signal that can be thought of as a  square wave  associated with the relativistic worldline. This introduces a discrete binary underlay to the worldline, representing an intrinsically discrete aspect of massive particles.  The signal itself reflects the fact that between any two  events is a causal spacetime {\em area} in 1+1 dimensions representing the intersection of the forward light cone of the first event and the backwards light cone of the second, \cfig{fig:incip}.
 
\begin{figure}[t]
\centering
\begin{subfigure}{.3\textwidth}
  \centering
  \includegraphics[width=.9\linewidth]{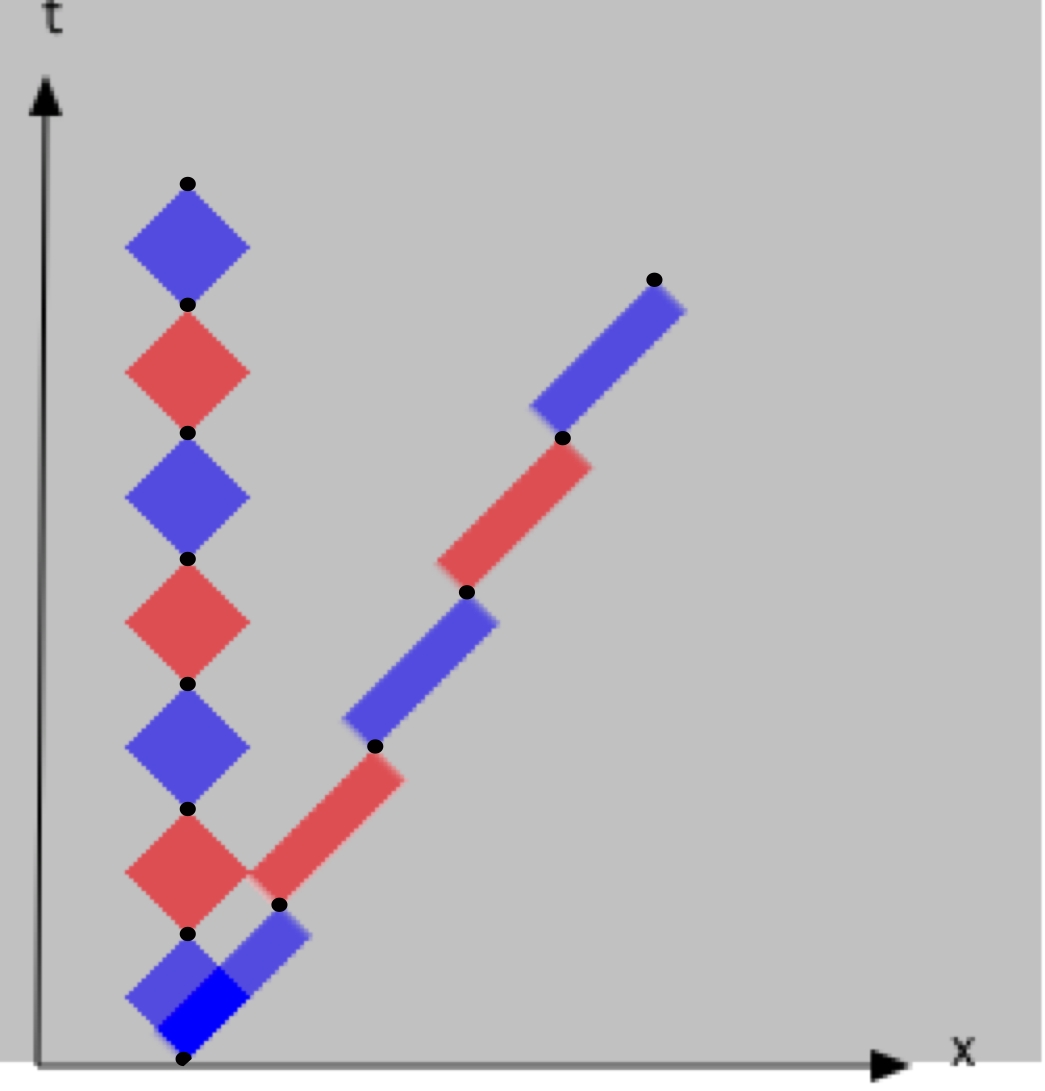}
  \caption{Periodic events join successive causal areas, with and without a boost.}
  \label{fig:sub1}
\end{subfigure}%
\hspace{2mm}
\begin{subfigure}{.3\textwidth}
  \centering
  \includegraphics[width=.9\linewidth]{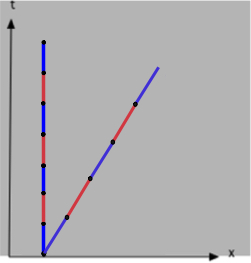}
  \caption{A projection of the area colouring onto the worldline gives a binary signal.}
  \label{fig:sub2}
\end{subfigure}
\hspace{2mm}
\begin{subfigure}{.3\textwidth}
  \centering
  \includegraphics[width=.9\linewidth]{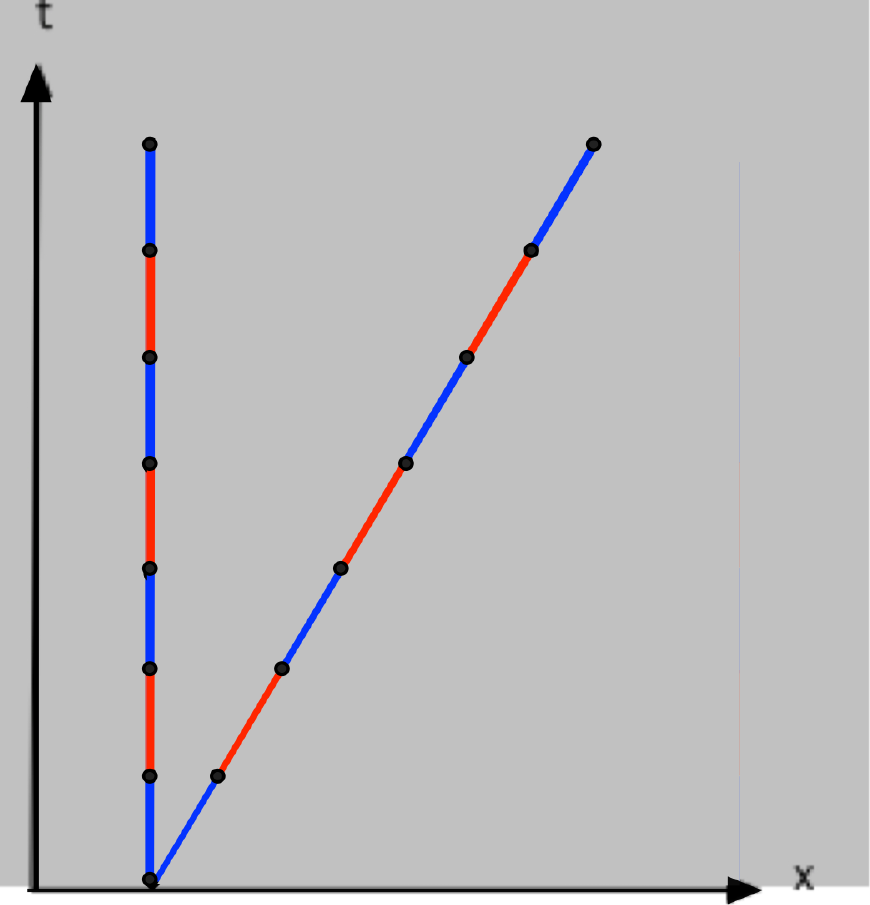}
  \caption{The Galilean transformation ignores time dilation, moving clocks stay synchronized.}
  \label{fig:sub3}
\end{subfigure}
\caption{Two C-particles starting in the same state leave the origin. One is stationary, one moving to the right. Here $c=1$ and the two colours of the causal areas between events display the binary aspect of the particle's signal. The events at the intersection of successive areas are considered single points that define the worldline on larger scales.  The binary aspect of the signal is generated by the causal areas that are successively distinguished by a single bit of information. Part (c) illustrates the fact that the Galilean transformation ignores time dilation.}
\label{fig:test}
\end{figure}

 For simplicity, we work here in units where $c=1$ and $m/\hbar$ is chosen to make the Compton wavelength  4. The nodes, maxima and minima of the `zitterbewegung' may then be chosen to occur at integers in the rest frame.  
 
 The binary aspect of the signal, referred to here as `parity', reflects a minimal variation
needed to mark time intervals, effectively establishing a clock with an intrinsic scale.  \cfig{fig:sub1} shows an image of a pair of clocks, one stationary  and one boosted, the colour differentiating successive intervals between events.  The Lorentz transformation giving the form of the boosted clock preserves the Euclidean area and colour of the causal areas, but in doing so stretches the period of the moving clock through time dilation.  \cfig{fig:sub2} shows the binary colouring of the worldline that results. For comparison \cfig{fig:sub3} shows the binary colouring of the worldline under the Galilean transformation where time dilation is absent.

 \begin{figure}[tbp] %  figure placement: here, top, bottom, or page
 \begin{minipage}{0.45\textwidth}
    \centering
    \includegraphics[scale=0.34]{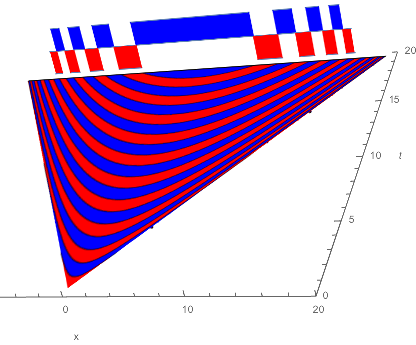} 
    \caption{The hyperbolae of constant proper time in the $x-t$ plane intersect a line of constant $t$. The wave pattern formed parallel to  the $x$-axis projects the clock's history from its maximal age at $(x,t)=(0,t)$ to the initial age at $(x,t)=(\pm t,t)$.  Particularly note the stretching of the `recent history' of the clock along the $x$-axis.}
    \label{fig:BoostClock}
    \end{minipage}\hspace{1cm}
     \begin{minipage}{0.44\textwidth}
    \centering
    \includegraphics[scale=0.46]{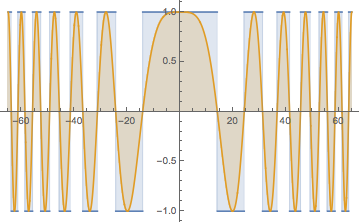} 
    \caption{The intersection of the hyperbolae of constant proper time at fixed $t$ from \cfig{fig:BoostClock}(light blue), plotted with the real part of the Feynman propagator(beige). This is for the central part of the pattern where the boost velocity $v<<1$. Here the binary clock is just the sign of the Feynman propagator.}
    \label{fig:VsFeyn}
 \end{minipage}   
 \end{figure}

If we indicate blue by $+1$ and red by $-1$, the coloured stationary signal illustrated in  \cfig{fig:sub2} can be written:
\bee
C_{0}(t)=\rm{sign}(\sin(\frac{\pi}{2} t))\label{eqn:1}
\ee
with the boosted clock with velocity $v$ giving
\bee
C_{v}(t)=\rm{sign}(\sin(\frac{\pi}{2} t \sqrt{1-v^{2}}))\label{eqn:2}
\ee
 \cfig{fig:BoostClock} shows the colouring of the $x-t$-plane from an ensemble of clocks in different inertial frames synchronized at the origin, the binary parity being distinguished by two colours. The characteristic hyperbolae of fixed proper time are evident. At the fixed value of $t=20$ in the figure, the parity of the boosted clocks is plotted using $+1$ for blue and $-1$ for red. As may be seen in the figure, the fixed $t$  signal is a representation of the clock's history that regresses to $t=0$ as $x$ approaches the light cone. This does not happen with the Galilean transformation which would {\em display a single colour at fixed $t$ regardless of $x$.}    

  In \cfig{fig:VsFeyn} an amplitude of the clock phase at fixed $t$ is plotted in comparison to the real part of the Feynman propagator for a non-relativistic particle of equal mass. For small relative velocities it is evident that 
    the binary clock has {\em the same sign and frequency as the propagator.} The clock signal that is a periodic square wave in time, \ceq{eqn:2}, results in a square wave of increasing frequency along the $x$-axis. It is worth noting that the broad maximum at the origin is in practice on the scale of the deBroglie wavelength $\hbar/mv$ rather than the Compton length.

 Although not plotted, the binary clock differs in frequency from the Feynman propagator near the light cone and goes to zero outside the light cone as would be expected. The Lorentz boost cannot take the worldline signal outside the future light cone.  In contrast, the Feynman propagator is not realistic near the light cone and continues oscillating for all $x$. This is not relativistically correct but is appropriate for the Schr\"{o}dinger equation with its infinite signal velocity.

 The binary clock `propagator' $C(x,t)$,  plotted in \cfig{fig:VsFeyn}  is a direct manifestation of time dilation in special relativity. The only input from quantum mechanics is the numerical value of the input frequency $mc^{2}/\hbar$. The result is however,  suggestive. From the figure, Feynman's propagator is a `softened' version of the binary signal that could arise from the erosion of the discontinuities by the introduction of  a stochastic element.  We shall explore this possibility in the next section.
 
  It is also apparent that in its present form, $C(x,t)$  squares to unity between $-t<x<t$ and thus $C^{2}(x,t)$ could be used as a probability density function at fixed $t$. The constancy  of $C^{2}$  suggests an interpretation  that all boost velocities would be equally weighted if $C(x,t)$ ultimately provided a probability density function.

 While the comparison  of $C(x,t)$ with wavefunctions and quantum mechanics  has   qualitative merit at this point, it is one thing to mimic a binary form of the phase of Feynman's propagator, it is quite another to mimic superposition. Special relativity is  ultimately a classical theory and the binary propagator would appear to be a classical object within that theory, its relation to the Feynman propagator notwithstanding. Superposition of wavefunctions rather than probability density functions is central to quantum interference and unless there is a specific reason that  binary `propagators' such as $C(x,t)$ rather than the probability density should add, the resemblance to quantum mechanics  remains a curious artifact.  
 
To probe the  question of superposition  for binary clocks, we consider an idealization of    a double slit experiment.\footnote{The double slit experiment is a typical choice to display the peculiarity of  `quantum' superposition because it highlights the {\em failure} of the classical superposition of probabilities. It also yields quickly to 'wave superposition' but is mute on the origin and physical reality of the waves.}  In order to do this, an extension of the inertial frame concept from special relativity to include `hinged' or `piecewise-inertial' frames is needed to consider clock signals traversing alternative paths. 
\begin{figure}[t!]
    \centering
    \begin{subfigure}[t]{0.45\textwidth}
        \centering
        \includegraphics[height=1.7in]{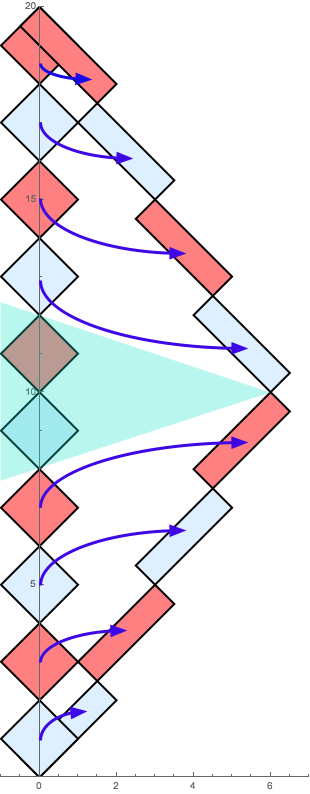}
        \caption{Here the hinged frame clock is a Lorentz boost image of the stationary clock both before and after the hinge. The small blue arrows illustrate the preimages of the hinged frame areas.}\label{fig:HingedSync}
    \end{subfigure}%
    \hspace{1cm}
    \begin{subfigure}[t]{0.45\textwidth}
        \centering
        \includegraphics[height=1.7in]{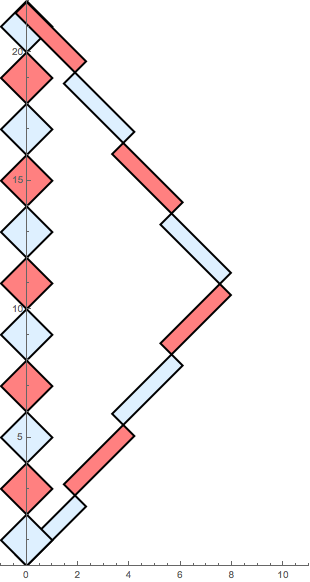}
        \caption{Here the hinged frame clock is not a Lorentz boost image of the stationary clock  after the hinge. As a result, the parity of the clocks disagree where they cross after the hinge.}\label{fig:HingedNoSync}
    \end{subfigure}
    \caption{An illustration of  stationary clocks compared to  hinged-frame clocks. A clock in a hinged frame instantaneously switches to a new inertial frame at an event, changing state as it does so.}
\end{figure}
 \noindent      \cfig{fig:HingedSync} shows an example of a hinged-frame clock. By hinged-frame clock we mean a clock  that instantaneously switches to 
another inertial frame {\em at an event}, changing state as it
does so. From the perspective of the `clock', the hinge is assumed to be an information handoff so that any physical effects of  acceleration in the velocity change is  hidden in an arbitrarily small spacetime region about an event.

   The hinged frame clock on the right of  \cfig{fig:HingedSync} may be thought of in two ways. By analogy with the `twin paradox' from special relativity, the hinged frame clock can be the `rocket twin';  a clock, identical to the rest frame clock, that happens to travel along the hinged frame path. Alternatively we can think of the hinged frame clock as simply {\em an image of the rest-frame clock under Lorentz boosts appropriate to the two frames.}  Before the hinge the boosted clock is {\em an image of the early history of the clock.} After the hinge the {\em image is of the late history. } The point of this interpretation is that in the figure, there are two images of the {\em same clock}.
   
    \begin{figure}[b!] %  figure placement: here, top, bottom, or page
    \centering
    \includegraphics[height=1.5in]{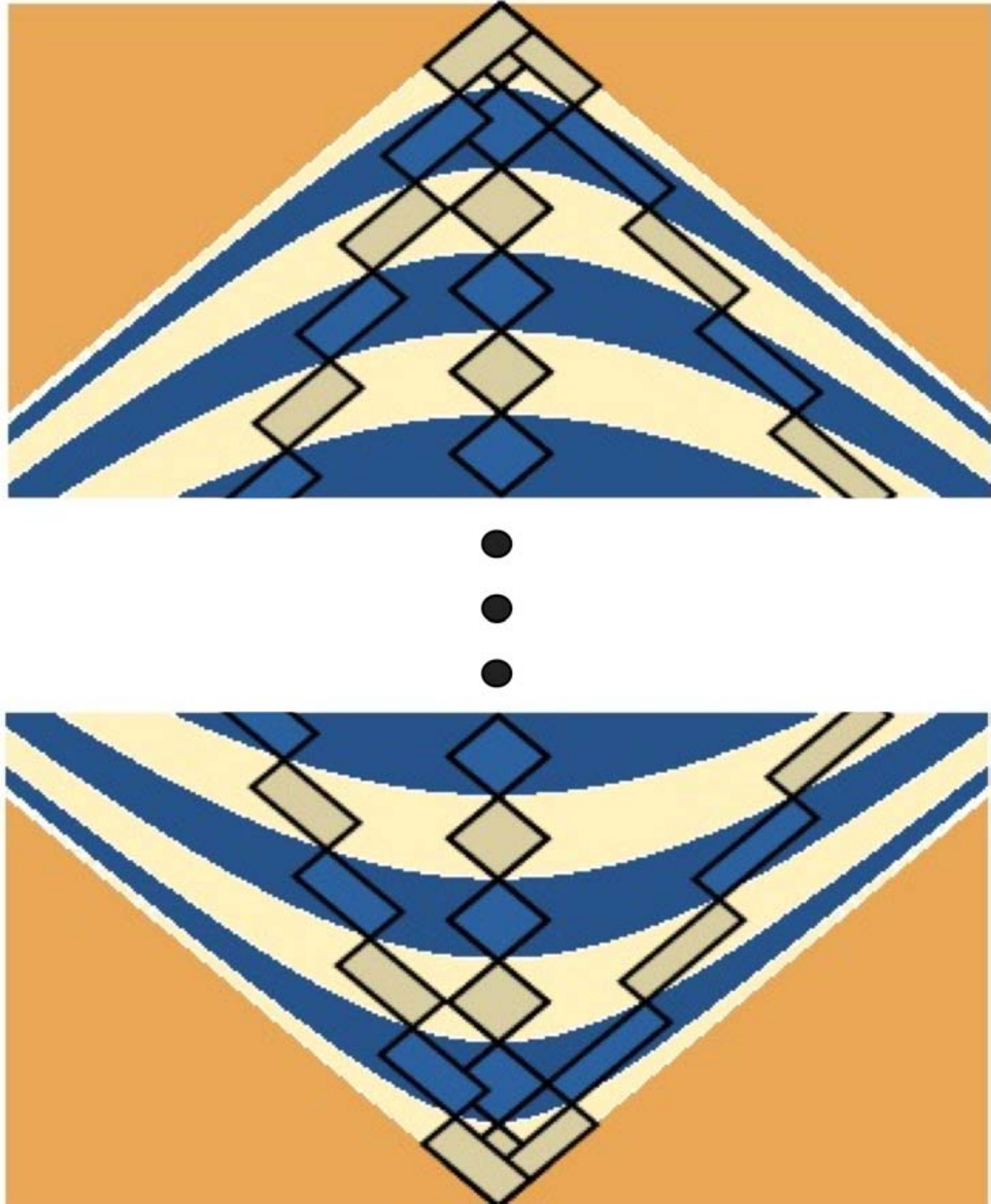} 
    \caption{Three paths between a source and a sink. The stationary clock has a well defined parity at the source and the sink. The two other implicitly hinged paths have the same parity at the source and sink. To be equivalent Lorentz images of each other, the hinged clocks must map onto each other at source and observation by omitting an integer number of full periods from the rest-frame clock.}
    \label{fig:SourceSink}
 \end{figure}

    In either interpretation, the hinged frame clock pictured differs from the rest frame clock in that it has one (or in general more)  full-period deletions of the rest frame clock. The full-period deletions ensure that the parity of the clocks agree where the paths cross.

 {\em In contrast} to the second interpretation of \cfig{fig:HingedSync} consider \cfig{fig:HingedNoSync}. Here the hinged-frame clock disagrees with the stationary clock with respect to parity at the end point. After the hinge, the hinged clock is {\em not} an image of the stationary clock under  Lorentz boosts. In this case the stationary and
  hinged clocks {\em must be distinct objects.} They cannot be simply a clock and its image in a hinged frame, or two images of the same clock.

 Let us apply hinged frame clocks to an analog of the double slit experiment. Assume that we send individual clock-particles through a double-slit apparatus and that each particle goes through one slit or the other with a randomly directed hinge at the exit of the slit. Assume the particle source is equidistant from the slits so the parity of the clock as it emerges from a slit will be the same, regardless of which slit it passes through. Now consider a point $x$ on the detector screen and  the two possible clock signals from the two slits, say A and B. If the signals A-$x$ and B-$x$ have the same parity at $x$ then, up to the binary discrimination of the clocks, the two hinged frames from the source to $x$ are Lorentz boosts of each other, before and after the hinge, as in \cfig{fig:HingedSync}. They can both be interpreted as images of the {\em same} C-particle signal from source to $x$. We call such pairs of paths Lorentz-equivalent, \cfig{fig:SourceSink}. 
 If on the other hand the signals A-$x$ and B-$x$ {\em disagree} in parity at $x$, then they are {\em not} both images of the same C-particle signal.  In this case,  we call such paths Lorentz-inequivalent.

 We are now in a position to question superposition in relation  to C-clocks. If we assume the binary clock parity labeling of $\pm 1$ given in \ceq{eqn:2} and we average the two possible binary clock signals from A and B at $x$ calling the result $\phi(x)$, then provided $x$ is in the light cone of both A and B we get:
 \begin{equation}
\phi(x)= \begin{cases}
  \pm 1    & \text{if paths are Lorentz-equivalent}, \\
   0  & \text{otherwise}.
\end{cases}
\label{eqn:filter}
 \end{equation}
Note that adding the binary signal here simply acts as a filter on the ensemble of paths from the source to the detector. It is conspicuous for what it filters out, namely, {\em those positions on the detector screen for which the two clock signals are not Lorentz images of each other.} Superimposing the signal rather than a probability density creates an ensemble of paths to the detectors that are Lorentz equivalent. In such a filtered ensemble, paths to a detector are all associated with a {\em single} clock signal. If we disallow the cancellation of path pairs by averaging the squares of the clock signal, we allow {\em into} the ensemble of paths histories of {\em  different} clocks. 

 \begin{figure}[t!] %  figure placement: here, top, bottom, or page
   \centering
   \includegraphics[scale=0.3]{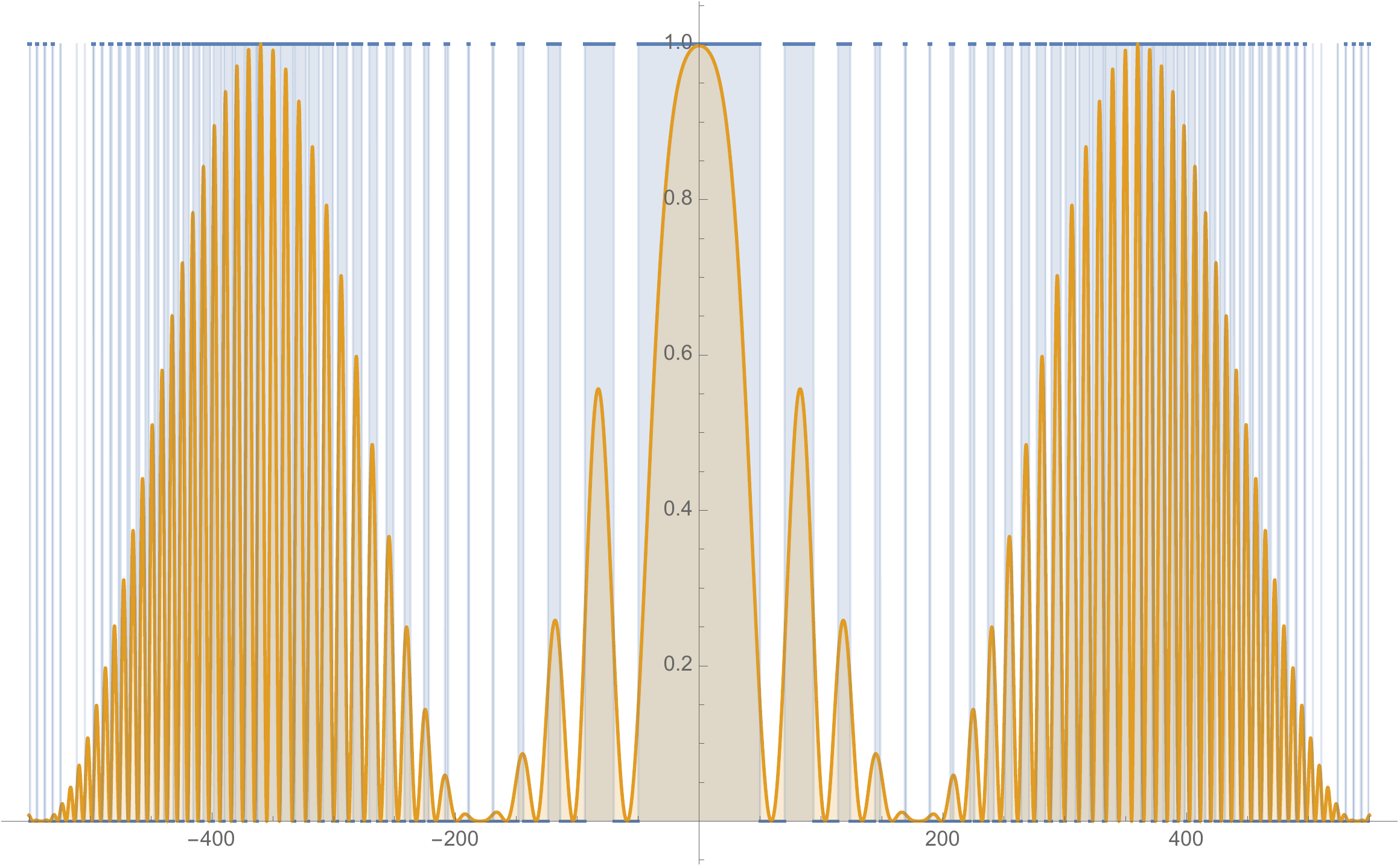} 
   \caption{The addition of the Feynman propagator from two spatially separated point sources shown at fixed t (smooth curve). The Real part shows the characteristic interference fringe with three maxima from a squared Cosine envelope restricting higher frequency components. Superimposed is the same result for hinged C-clocks that arrive at the same place. The binary clock  amplitude squared $\phi^{2}(x)$ ( see eqn. \ref{eqn:filter}), coloured light blue  for the same double source is shown. The gaps in $\phi^2(x)$ occur where paths are Lorentz inequivalent and the binary choice reflects the filtering of \ceq{eqn:filter}. The results  imitate interference fringes produced by the superposition of  the path integral amplitudes from the two slits. However, $\phi^{2}(x)$ is a direct result of special relativity and would be constant in its absence. }
   \label{fig:nodes}
\end{figure}

 If we then square $\phi(x)$ we just get a constant function with gaps where \mbox{ $\phi^{2}(x)=0$, } the gaps occurring where paths are inequivalent, \cfig{fig:nodes}. $\phi^{2}(x) \in \{0, 1\}$ could then be used as a probability density function for the arrival of C-clocks, the gaps representing nodes where arrival is prohibited. Notice that in \cfig{fig:nodes}, the Feynman propagator inherits its superposition principle from the Schr\"{o}dinger equation for which it provides a path-integral formulation. Its connection to probability density functions is likewise through the Born postulate, an \mbox{a posteriori }result verified by experimental evidence.   By comparison, the addition of two signals by  $\phi(x)$ has an obvious interpretation as a binary Lorentz filter.  It has an a priori function as a coarse sieve that accepts pairs of paths if they agree up to parity at source and detector, and rejects them otherwise. The connection to probability density functions is then potentially as direct as the classical case. Squaring $\phi(x)$ gives a measure of allowable paths through the two slits given Lorentz filtration. {\em Passage of a C-clock through one or other slit are no longer disjoint events given that detection involves a restriction on parity.}

That the superposition of binary clock signals is the source of `interference' in $\phi(x)$ is unambiguous. If we superimposed the square of the signals as appropriate for the classical addition of probability densities, we would arrive at the uniform distribution where the light cones for the two slits intersect. There would be no gaps in the probability density function.

Taking into account  that superposition of C-clock signals eliminate Lorentz-inequivalent paths, giving interference effects similar to those  of quantum mechanics, the question arises as to whether the Schr\"{o}dinger and Dirac equations emerge from special relativity in the same way. In terms of the latter equation, the Feynman chessboard model is a path-integral interpretation of a `sum-over-hinged-frames' using exactly the same parity device to eliminate Lorentz-inequivalent paths. This is implicit in  the chessboard model itself \cite{gaveau:419,jacschul84,Kauffman:1996kq,GNORBM2011} and will be made explicit in a subsequent publication. 

The origin of the Schr\"{o}dinger equation in terms of the filter in eqn(\ref{eqn:filter}) follows from the chessboard model in the non-relativistic limit. There are subtleties to this limit but  the effect of (\ref{eqn:filter}) by a counting of parity in a simple four-state `clock' has been verified\cite{coexist,gordDeakin96} in a model that we sketch below in the context of hinged frames. 

\section{A Stochastic Model}In the clock-particle case discussed above, parity  keeps track of elapsed time by counting the number of `corners' in the causal areas between events. This makes sense relativistically in that the  segments between corners are null and do not evolve proper time. Parity then distinguishes between an even and odd number of direction changes in the causal envelopes of paths \cfig{fig:sub1}. We can preserve this property in a non-relativistic model by taking a `diffusive' limit. For large time and space scales, the association of time evolution with direction change is necessarily unrealistic since individual steps in the random walk are covered at small speeds much less than $c$. However, in the diffusive limit, the mean free speed  becomes arbitrarily large as space and time steps become small, making the approximation of time evolution with direction change a good one on small scales, once the mean free speed exceeds $c$. Thus,  a simple model of diffusion  keeping track of successive direction changes in paths, using the number of such changes to define parity, approximates the relativistic feature that inter-event paths are null.

To implement this, following \cite{coexist,gordDeakin96},   define densities 
 $p_\mu(m \delta, s \epsilon)$, 
$(\mu = 1,2,3,4)$. These  represent the probabilities
that a C-particle leave a
space time point $(m \delta, s \epsilon)$  in state $\mu$
$(m= 0,\pm 1,\ldots;\;s=0,1,\dots)$.
The difference equations for $p_\mu$  are
$$\begin{array}{rcl}
p_1(m\delta,(s+1)\epsilon) & = & \frac{1}{2} p_1 ((m-1)\delta,s \epsilon)
 + \frac{1}{2}
p_4((m+1)\delta, s \epsilon) \\
p_2(m\delta,(s+1)\epsilon) & = & \frac{1}{2} p_2 ((m+1)\delta,s
\epsilon) + \frac{1}{2} p_1((m-
1)\delta, s \epsilon) \\
p_3(m\delta,(s+1)\epsilon) & = & \frac{1}{2} p_3 ((m-1)\delta,s
\epsilon) + \frac{1}{2}
p_2((m+1)\delta, s \epsilon) \\
p_4(m\delta,(s+1)\epsilon) & = & \frac{1}{2} p_4 ((m+1)\delta,s
\epsilon) + \frac{1}{2} p_3((m-
1)\delta, s \epsilon).
\end{array} \eqno(2) $$
\addtocounter{equation}{1}

These equations express the conservation of the number of particles over time, with half of them maintaining their direction and state at a time step, and half changing direction and state. The four states refer to the two spatial directions and the two possible values of parity.

To express (2) in matrix form, consider the shift operators
$E_x^{\pm1}$ and $E_t$ such that
\begin{eqnarray}\nonumber
E_x^{\pm1} \; p_i(m \delta, s \epsilon) &=& p_i((m \pm 1)\delta, s
\epsilon)\quad  {\rm and}\\ \nonumber
E_t \quad p_i(m \delta, s \epsilon) &=&p_i(m\delta, (s+1)
\epsilon).\end{eqnarray}
The difference equations (2) may then be written as
$$E_t \; P(m \delta, s \epsilon) = \frac{1}{2}
\left[\begin{array}{cccc}E_x^{-1}&0&0&E_x\\
E_x^{-1}&E_x&0&0\\
0&E_x&E_x^{-1}&0\\
0&0&E_x^{-1}&E_x \end{array}\right] P(m
\delta, s\epsilon) \eqno(3)$$
\addtocounter{equation}{1}
\noindent where
$P(m \delta, s \epsilon) $ is a column vector of the $p_\mu$. Now consider a change of variables:

\begin{equation}
z_1 =\frac{p_1 +p_3}{2},\;z_2 =\frac{p_2+p_4}{2}\label{eqn:cofz}
\end{equation}

and
\begin{equation}
 \phi_1 =\frac{p_1 -p_3}{2},\;\phi_2 =\frac{p_2 -p_4}{2}.\label{eqn:cofv}
\end{equation}

The $z_{k}$  just represent probabilities, partitioned by direction. The $\phi_{k}$ record {\em parity} in the system, partitioned by direction. In terms of counting paths, the $\phi_{k}$ record the net number of paths that are Lorentz equivalent using the $\pm 1$ filtering process of the C-clock signal. Eqn(\ref{eqn:cofv}) is the implementation of eqn( \ref{eqn:filter}) in this model.

 The change of variables block diagonalizes the shift matrix to give:
\begin{equation}
E_t \left[\begin{array}{c}z_1\\ z_2\\ \phi_1\\
\phi_2 \end{array}\right] =
\frac{1}{2} \left[\begin{array}{cccc}E_x^{-1}  &E_x
 &0&0\\
E_x^{-1}&E_x&0&0\\
0&0&E_x^{-1}&-E_x\\
0&0&E_x^{-
1}&E_x\end{array}\right]\;\left[\begin{array}{c}z_1\\ z_2\\
\phi_1\\ \phi_2
\end{array}\right].\end{equation}
The upper block gives a discrete form of the diffusion equation,

\begin{equation}
E_t \left[\begin{array}{c} z_1\\
z_2 \end{array}\right] =
\frac{1}{2} \left[\begin{array}{cc}E_x^{-1}&E_x\\
E_x^{-
1}&E_x\end{array}\right]\;\left[\begin{array}{c}
z_1\\ z_2
\end{array}\right] 
\label{eqn:shift}\end{equation}  the lower block is:
\begin{equation}
E_t \left[\begin{array}{c} \phi_1\\
\phi_2 \end{array}\right] =
\frac{\alpha}{2} \left[\begin{array}{cc}E_x^{-1}&-E_x\\
E_x^{-
1}&E_x\end{array}\right]\;\left[\begin{array}{c}
\phi_1\\ \phi_2
\end{array}\right] 
\label{eqn:shift}\end{equation} where  a normalization constant $\alpha$ has been inserted\footnote{Note in particular the similarity of the two equations. The difference lies in the off-diagonal elements of the spatial operator. To lowest order the shift operators are unity. In (11), to lowest order the off-diagonal matrix is $\mathbf{\sigma}_{x}$ with eigenvalues $\pm 1$, indicating a reflection. In (12) it is $-i \mathbf{\sigma}_{y}$ with eigenvalues $\pm i$, indicating a rotation!}. The constant is necessary to allow  the filtered paths in the continuum limit to survive diffusive scaling. If we keep $\alpha=1$  the parity filtered ensemble is dominated by the full ensemble of diffusive paths and the effect of these paths  in relation to all diffusive paths will be lost in the continuum limit.

Consider now the generating function (discrete Fourier transform)
\begin{equation}\phi_k(p,s\epsilon) =\sum_{m=-\infty}^\infty \phi_k(m\delta,s\epsilon)e^{-ipm\delta}\end{equation}
Using eqn(\ref{eqn:shift}) the shift in time is
\begin{equation}
\begin{pmatrix}
\phi_1(p,(s+1)\epsilon)\\
\phi_2(p,(s+1)\epsilon)
\end{pmatrix}
=T \begin{pmatrix}
\phi_1(p,s\epsilon)\\
\phi_2(p,s\epsilon)
\end{pmatrix}=
T^{s+1} \begin{pmatrix}
\phi_1(p,0)\\
\phi_2(p,0)
\end{pmatrix}
\end{equation}
where $T=\frac{\alpha}{2} \begin{pmatrix}
e^{-i p \delta}&-e^{i p \delta} \\
e^{-i p \delta}&e^{i p \delta} \\
\end{pmatrix}
$ is the transfer matrix. To take a continuum limit  large powers of $T$ are needed. The eigenvalues of $T$ are

 \begin{equation}
 \lambda_\pm =\frac{\alpha}{\sqrt{2}} e^{\pm i \pi/4} \left (1 \pm i \frac{p^2 \delta^2}{2} +O(\delta^4)\right ).
 \end{equation}
 
 To extract a continuum limit it is necessary to choose $\alpha =\sqrt{2}$ and to make sure the powers are taken through a sequence of integers that are 0 mod 8. This is because each step in the process advances the state of  half the clocks giving 8 as the expected number of steps to a return to the original state. Removing the fine-scale state changes with a stroboscopic limit removes the 'zitterbewegung' that is associated with the relativistic
clock, allowing the larger scale pattern to emerge.  Considering the usual diffusive limit:
\begin{equation}
\{\delta \to 0,\; \epsilon \to 0,\; \frac{\delta^2}{\epsilon}\to 2D,\; m\delta\to x,\; s\epsilon \to t \}
\end{equation}
 with the mod 8 restriction applied, $\lim_{\delta\to 0}\lambda^s_\pm =e^{\pm i p^2 D t}$  and the propagator is
\begin{equation}
\lim_{s\to \infty}\Phi(p, s\epsilon)=\begin{pmatrix}\cos(p^2 D t)&-\sin(p^2 D t)\\ \sin(p^2 D t)&\cos(p^2 D t)\\\end{pmatrix} \Phi(p, 0)
\end{equation}
To find  a more familiar form, write
\begin{eqnarray}\label{eqn:iin}
\psi_+(p,t)&=&\left (\; i \phi_1(p,t) + \phi_2(p,t)\, \right)/2\\\nonumber
\psi_-(p,t)&=&(\, -i \phi_1(p,t) + \phi_2(p,t)\,)/2,\\\nonumber
\end{eqnarray}
\vspace{-9mm} 

\noindent take $\psi_{\pm}(p,0)=\frac{1}{\sqrt{2}}$ and transform back to position space to give 

\begin{equation}
\Psi(x,t)=\begin{pmatrix}\frac{e^{ix^2/4Dt}}{\sqrt{4 \pi i D t}}&0\\0&\frac{e^{-ix^2/4Dt}}{\sqrt{-4 \pi i D t}} \\ \end{pmatrix}\frac{1}{\sqrt{2}}\begin{pmatrix}1\\1 \\ \end{pmatrix}.\label{eqn:fin}
\end{equation}
 Here, it is apparent that the two  components of $\Psi$ satisfy conjugate Schr\"{o}dinger equations. Notice it is the association of $\pm 1$ as the parity giving the definitions of the $\phi_{k}$ in (\ref{eqn:cofv}) that extracted the wave propagator, just as it was the use of binary discrimination, \ceq{eqn:filter}, that imitated interference in the first section.  It is also worthwhile noting that the use of the unit imaginary $i$ in \ceq{eqn:iin}  is not a formal analytic continuation, it is simply a convenience to bring the propagator to a familiar form. An application of the procedure from (7) to (12)  to the $z_{k}$  with $\alpha=1$ produces the Green's function for  the diffusion equation\cite{gordDeakin96}\footnote{In the case of the $z_{k}$, $\alpha =1$ and the eigenvalues of the transfer matrix are 0 and $\cos( p \delta)$. In the diffusive limit, after transformation back to position space the analog of \ceq{eqn:fin} is $Z(x,t)=\frac{1}{2}\left(\begin{array}{c}1 \\1\end{array}\right) \frac{1}{\sqrt{4 \pi D t}}e^{-x^2/4Dt}$, the diffusive Green's function. Notice that the formal analytic continuation that takes the diffusion equation to the Schr\"{o}dinger equation is in this context no longer formal but specified by keeping track of parity through \ceq{eqn:cofv}.}.
 
  \section{Discussion}
  
  Looking through the above derivation it is apparent that special relativity and aspects of quantum propagation are  deeply connected. On one hand,  replacement of the binary C-clock signal by the constant function, reinstating the conventional scale-free form of a worldline, removes the binary phase that drives the `quantum' superposition principle. Classical special relativity with scale-free worldlines would be  recovered as a result. 
  
  Conversely, if  the classical worldline of special relativity is marked by a binary periodic signal and  Lorentz boosts of those signals over hinged frames  are filtered to agree in parity, then the Schr\"{o}dinger equation and its appropriate superposition principle emerge. {\em Conceptually, special relativity allows us to toggle between the classical theory and quantum propagation by the simple expedient of switching between a scale-free and a binary periodic version of the worldline concept.}
  
  Returning to the argument (\ref{quote:arg}) favouring the view that RQM and its derivatives are {\em effectively} extensions of quantum mechanics, we can now appreciate the weakness of this position. Time dilation is completely absent from Newtonian physics, yet here it {\em directly implicates} the Schr\"{o}dinger equation. We can see how time dilation survives the implicit $c\to \infty$ limit by looking at \cfig{fig:BoostClock} and noticing that the Lorentz transformation acts as a magnifying glass  focussed on the recent history of the clock. The very high frequency of the zitterbewegung associated with $m_{0}c^{2}$ is removed in the non-relativistic limit, however  the Lorentz transformation stretches the zitterbewegung at the Compton frequency to produce the relatively slowly varying phase at the deBroglie scale {\em without the explicit appearance of the speed $c$.}\footnote{The failure of $c$ to appear explicitly is analogous to the failure of $c$ to appear in the $O(v^{2})$ term in the  expansion of $mc^{2}$. Phase in Schr\"{o}dinger wavefunctions are not overtly linked to special relativity by an association with $c$ for the same reason.} This associates an intrinsic signal with mass and momentum. The equivalence of inertial frames provides an ensemble of equivalent signals thereby making the Fourier uncertainty principle `physical'  and tied to the Lorentz transformation.

  From the perspective of the clock model, the utility of the wavefunction in NRQM is to introduce a form of relativistic filtering that is absent from the classical mechanics conventionally underlying the Schr\"{o}dinger equation.  Probing this equation and the underlying classical mechanics without considering special relativity may uncover interesting relationships, but it is unlikely to uncover the superposition principle as an emergent feature. However, from the above model, stepping back to discrete events in a relativistic context allows emergence of superposition.
  
  \section{Conclusions}
    The argument that relativistic quantum mechanics is effectively an extension of non-relativistic quantum mechanics is supported by the close association between  Hamiltonian mechanics and the Schr\"{o}dinger's equation.  However, both the history and pedagogy surrounding quantum mechanics implicitly assume a stronger result, that in fact the quantum phenomena described by Schr\"{o}dinger's equation exist independently of special relativity.  The fact that non-relativistic quantum mechanics is self-contained is commonly taken as evidence that the phenomena it describes would exist in a world where special relativity was not present.
    
    The C-model above shows that this view is unlikely to be correct. Quantum mechanics aside, the transition from relativistic to Newtonian mechanics works effectively because classical particles in collisions at non-relativistic speeds conserve, to a good approximation, non-relativistic momenta, energy {\em and rest mass}. This allows a benign dismissal of both rest energies and high order terms in $v^{2}/c^{2}$ giving a self-contained `Newtonian Mechanics'.   However, we have seen that in the case of a discrete inner scale, discussed above, the worldline of a particle becomes a signal that must be constrained `mathematically' by the Fourier uncertainty principle, and `physically' by the equivalence principle.
    
    How these two principles are resolved depends on how and when the continuum limit is taken. The C-clock shows that if the continuum limit is taken last, while enforcing a `low-pass' filter to remove zitterbewegung, the result is the Feynman propagator, but with insight into the origin and role of phase.  If the continuum limit is taken first, the starting point becomes a set of partial differential equations but the provenance of these equations is lost!
    
The emergence of both phase and superposition from the C-clock model suggests that quantum mechanics may well be an  intrinsically relativistic effect, special relativity providing the scaffolding upon which quantum mechanics is built. The absence of  `$c$' notwithstanding,  time dilation lurks beneath the non-relativistic veneer of Schr\"{o}dinger's equation.

\section{References}
 \bibliographystyle{unsrt}

\end{document}